\documentclass[dvips]{article}
\usepackage{spie}
\usepackage{amsmath}

\input{psfig.sty}   

\title{Spectral characteristics of ultrashort pulses in Kerr-lens mode-locked lasers} 

\author{V.L. Kalashnikov, E. Sorokin, and I.T. Sorokina
\skiplinehalf 
Institut f\"ur Photonik, TU Wien, Gusshausstr. 27/387, A-1040 Vienna, Austria}

\authorinfo{Further author information: (Send correspondence to Dr. E. Sorokin)\\E. Sorokin: E-mail: e.sorokin@tuwien.ac.at}
 
\begin{document} 
  \maketitle 

%%%%%%%%%%%%%%%%%%%%%%%%%%%%%%%%%%%%%%%%%%%%%%%%%%%%%%%%%%%%% 
\begin{abstract}
A number of factors that influence spectral position of the femtosecond pulse
in a Kerr-lens modelocked Cr:LiSGaF laser have been identified: high-order
dispersion, gain saturation, reabsorption from the ground state, and
stimulated Raman scattering. Using the one-dimensional numerical model for
the simulation of the laser cavity, the relative contributions of different
factors have been compared. The Raman effect provides the largest
self-frequency shift from the gain peak (up to 60 nm), followed by the gain
saturation ($\sim 25$ nm), while the high-order dispersion contribution is
insignificant ($\sim 5$ nm). Comparison with the experimental data confirm
that the stimulated Raman scattering is a main cause of the ultrashort pulse
self-frequency shift observed in Cr:LiSGaF and Cr:LiSAF lasers.
\end{abstract}


%%%%%%%%%%%%%%%%%%%%%%%%%%%%%%%%%%%%%%%%%%%%%%%%%%%%%%%%%%%%%
\section{INTRODUCTION}
\label{sect:intro}  

A rapid progress in Kerr-lens mode-locking technique allows to
reach 14 and 12-fs pulse durations in Cr:LiSGaF and Cr:LiSAF
lasers, respectively ~\cite{14fs,12fs}. These active media are
attractive due to possibilities of sub-20 fs pulse generation
directly from the diode-pumped laser. At the same time, the lasers
demonstrate a significant Stokes shift of the pulse spectrum at
such short pulsewidths ~\cite{sub20fs,ICONO,14fs}. This shift
decreases the accessible bandwidth due to the worse overlap of the
gain and the pulse spectra, thus setting a limit to pulse duration
~\cite{ICONO}. Therefore, investigation of the nature of the
ultrashort pulse spectrum transformation has not only academic but
also practical significance.

A number of explanations for the ultrashort pulse spectrum shift
in mode-locked lasers have been suggested. For example, it was
supposed, that due to the high-order dispersions the spectral
region of negative group-delay dispersion, which is optimal for
pulse formation, may be displaced ~\cite{UBBFL}. However in the
framework of the perturbation theory the third-order dispersion
does not contribute to the pulse carrier frequency, but results in
spectrum asymmetry and fragmentation ~\cite{Akhmanov}. Only in the
case of the non-steady-state operation there is a possibility for
the strong dynamic frequency shift ~\cite{NewMex}.

A more realistic explanation of the frequency shift takes into account
frequency filtering due to reabsorption in the gain medium ~\cite{12fs}.
However, as it will be shown below, this explanation can not explain the
dependence of the frequency shift on the pulse energy. Moreover, such
dependence suggests that a nonlinear mechanism is involved in the frequency
shift.

As it was shown in Ref.~\cite{JosaHaus}, the stimulated Raman
scattering in active medium is a very suitable nonlinear process,
which can produce the experimentally observed Stokes frequency
shift in sub-50 fs domain. The analysis in Ref.~\cite{JosaHaus}
was based on the soliton perturbation theory that did not allow to
take into account the effect of high-order dispersion and
frequency-dependent dissipative lasing factors such as asymmetric
gain band, spectral filtering on the output mirror, reabsorption
in the gain medium and gain saturation.

In this work we performed a numerical analysis of the spectral
characteristics of ultrashort pulses on the basis of a relatively simple
model of the Kerr-lens mode locking (KLM). The main advantages of our model
is taking into consideration of the high-order dispersion, exact profiles of
the loss and gain bands, frequency dependent reabsorption in the active
medium, gain saturation and fast absorber action of the Kerr-lensing. The
obtained results are in a good agreement with experimental data and allow to
estimate the contribution of the different factors to spectral
characteristics of ultrashort pulses.

The paper is organized as follows: first a summary of relevant measurements
and experimental results is given, followed by the construction of the
analytical and computational models. We then present the results of our
calculations and discuss the influence of each of the abovementioned factors
separately. Finally, we present the results of simulation with all factors
included, using distributed and discrete-element models.

%%%%%%%%%%%%%%%%%%%%%%%%%%%%%%%%%%%%%%%%%%%%%%%%%%%%%%%%%%%%%
  \section{EXPERIMENTAL OBSERVATIONS} 

\begin{figure}
   \begin{center}
   \begin{tabular}{c}
   \psfig{figure=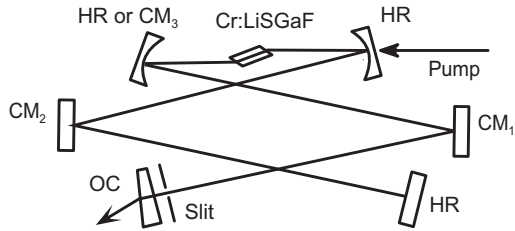,height=3cm} 
   \end{tabular}
   \end{center}
   \caption[example] 
   { \label{setup}	  
General scheme of a KLM Cr:LiSGaF laser used in this paper.
This scheme directly corresponds to the experiments in Refs~\cite{14fs,sub20fs}. HR, high reflector. CM, chirped mirror. OC, output
coupler.} 
   \end{figure}

\begin{figure}
   \begin{center}
   \begin{tabular}{c}
   \psfig{figure=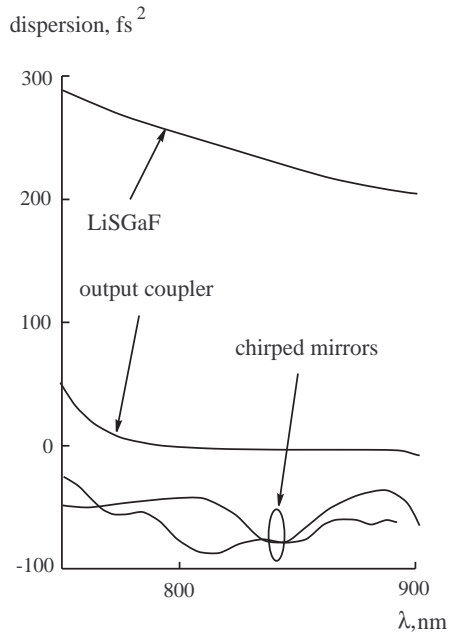,height=9cm} 
   \end{tabular}
   \end{center}
   \caption[example] 
   { \label{DispFig}	  
Measured group delay dispersion of the  active media (8
mm in double pass), output coupler and chirped mirrors in dependence on the
wavelength $\lambda$.} 
   \end{figure}

\begin{figure}
   \begin{center}
   \begin{tabular}{c}
   \psfig{figure=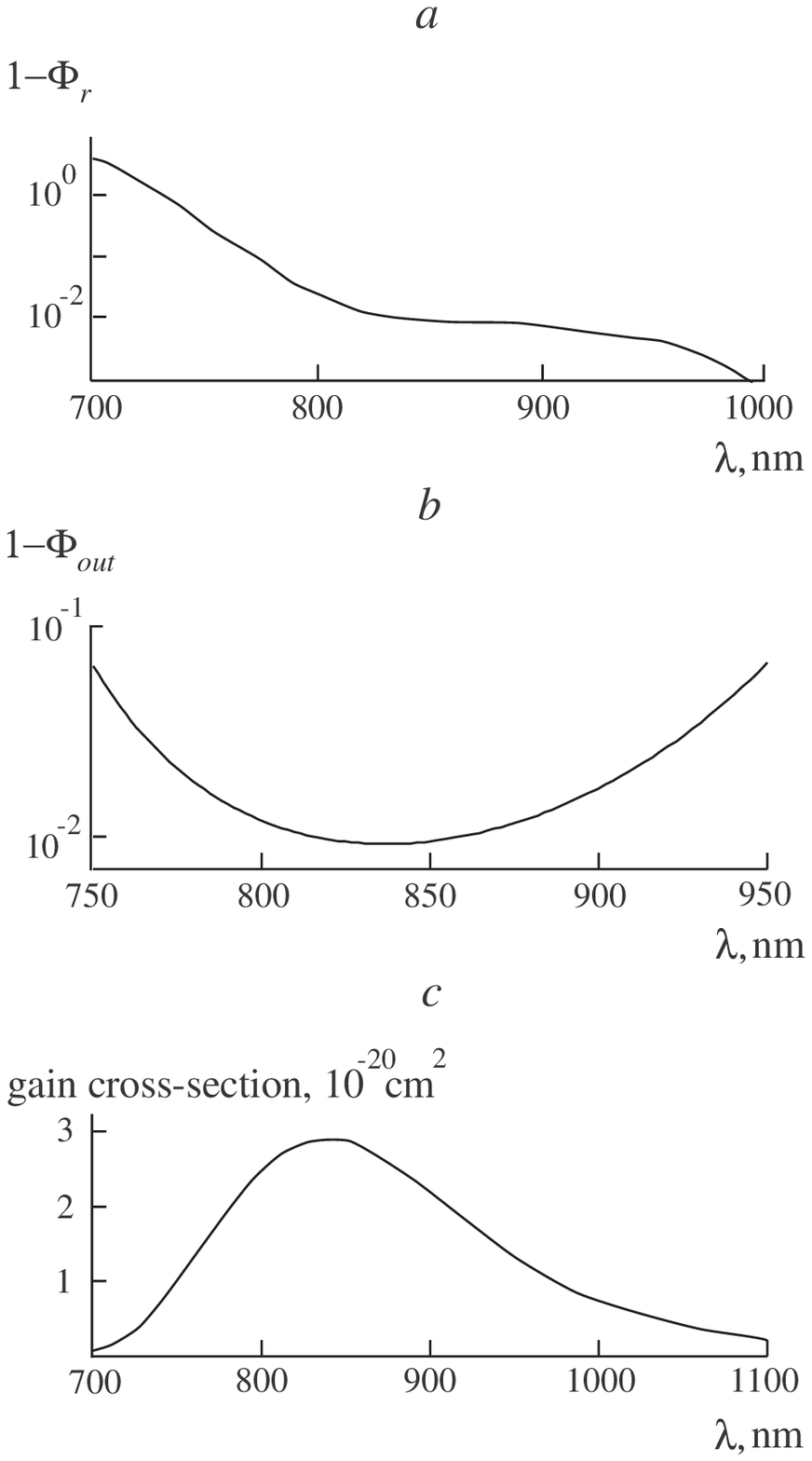,height=12cm} 
   \end{tabular}
   \end{center}
   \caption[example] 
   { \label{LossGain}	  
Ground-state absorption of Cr:LiSGaF (\textit{a}),
round-trip resonator losses due to the output coupler (\textit{b}) and gain
cross-section of Cr:LiSGaF (\textit{c}).} 
   \end{figure}

Systematic Stokes shift of the ultrashort modelocked pulse in Cr:LiSGaF- and
Cr:LiSAF-based KLM oscillators has been first reported in 1997
~\cite{14fs,sub20fs,TOPS98}, where the stimulated Raman scattering has been
suggested as a possible mechanism. It has been found, that i) the shift could
be observed in oscillators with different dispersion characteristics, ii) the
shift increases with the pulse energy, and iii) the shift increases with
pulse shortening. At pulse durations below 20 fs, the peak of the pulse
spectrum may shift as far as 50-70 nm into the infrared as compared to the cw
wavelength or modelocked spectrum at long pulse durations ($\sim $840 nm in
Cr:LiSGaF) ~\cite{14fs}. Later, analogous shifts in Cr:LiSAF have been
reported by Uemura and Torizuka ~\cite{12fs} and R. G\"abel \emph{et al}
~\cite{Kai}. All mentioned experiments used the common optical scheme,
differing only in pump arrangements and dispersion compensation techniques.
The schematic diagram of the laser oscillator is shown in Fig.~\ref{setup}.
This is a representative scheme for any X-shaped KLM laser, because different
types of dispersion compensation can always be represented by lumped
dispersion of a chirped mirror. For modeling purposes we used the parameters
of experiments, reported in Refs.~\cite{14fs,TOPS98}: the Brewster-angled 4
mm long LiSGaF crystal doped with 1.4\% Cr, beam diameter in Cr:LiSGaF
crystal 40$\times$60~$\mu$m. The high reflectors (HR) had negligible
dispersion. The dispersion of the chirped mirrors (CM) has been calculated
from its original design, and additionally measured by white-light
interferometry. The intracavity dispersion was calculated using the
dispersion data of Cr:LiSGaF ~\cite{sub20fs}, measured dispersion curves of
the chirped mirrors and calculated dispersion of the output coupler
(Fig.~\ref{DispFig}). We also used the experimental loss spectra (mirror
transmission and ground-state absorption in Cr:LiSGaF) as shown in
Fig.~\ref{LossGain}.

%-------------
%-------------

%-------------
%-------------

%-------------
%-------------

\begin{figure}
   \begin{center}
   \begin{tabular}{c}
   \psfig{figure=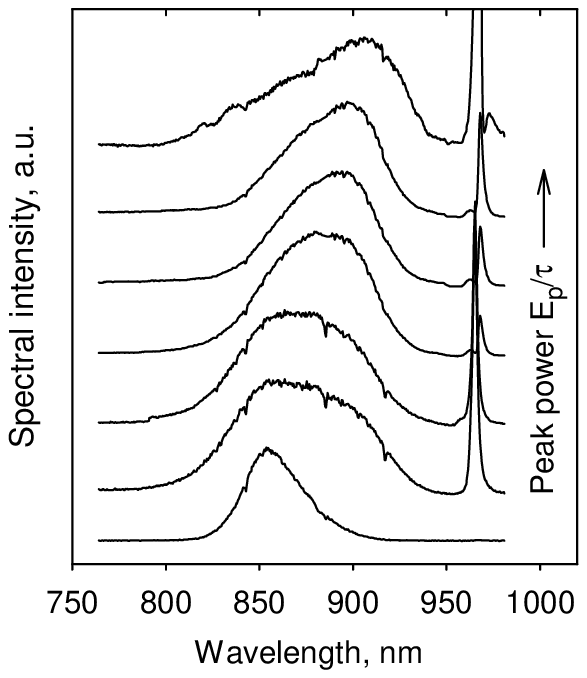,height=7cm} 
   \end{tabular}
   \end{center}
   \caption[example] 
   { \label{exshift}	  
Experimental demonstration of the pulse frequency shift.
The output spectra of modelocked Cr:LiSGaF laser, with changing intracavity
pulse peak power (bottom spectrum corresponds to the lowest peak power, top
spectrum - to the highest peak power).} 
   \end{figure}

The laser was pumped by 1.2-1.5 W from a Kr$^{+}$-ion laser at 647~nm, with a
TEM$_{00}$ beam, of which 0.9-1.1 W have been absorbed in the active medium,
generating 60-100 mW of average output power in the modelocked regime. The
resonator was slightly asymmetric, with distances between the curved mirrors
(radii of curvature 100~mm) and end mirrors being 88 and 109~cm,
corresponding to 72~MHz pulse repetition rate. Modelocking has been achieved
primarily by the hard aperture in form of an adjustable slit close to the
output coupler (Fig.~\ref{setup}). Fig.~\ref{exshift} shows normalized output
spectra at different pulse peak power, demonstrating the spectral shift.
Similar spectral behaviour has been observed also in Cr:LiSAF oscillators.
However, for the sake of simplicity we provide experimental data and perform
simulations on Cr:LiSGaF only.

\begin{figure}
   \begin{center}
   \begin{tabular}{c}
   \psfig{figure=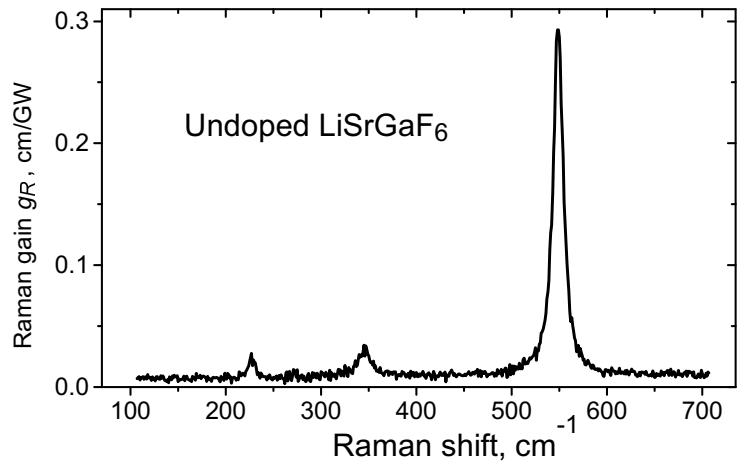,height=7cm} 
   \end{tabular}
   \end{center}
   \caption[example] 
   { \label{lgraman}	  
Raman gain of undoped LiSGaF. Exciting laser line at
514.5 nm and scattered light are both polarized along the crystallographic
$z$ axis, corresponding to the polarization of the laser radiation in
Cr:LiSGaF laser.} 
   \end{figure}

The Raman gain spectrum of undoped LiSGaF crystal has been
measured according to the procedure described in
Ref.~\cite{TOPS98}, using the orientated crystalline quartz as a
reference, and taking into account the thermal phonon population
factor. The spectrum (see Fig.~\ref{lgraman}) is obtained from
spontaneous Raman scattering spectrum, recordered in confocal
back-scattering geometry. Both incident and scattered light is
polarized along the $c$ axis, corresponding to the polarization of
light in the laser. LiSGaF possesses spatial symmetry group
D$^2_{3d}$ (P$\Bar{3}$1c) with 2 formula units in a unit cell,
resulting in total of 32 optical phonon modes
$3A_{1g}+4A_{2g}+4A_{1u}+5A_{2u}+8E_g+8E_u$, of which
$3A_{1g}+8E_g$ are Raman-active. In the scattering geometry as
described above, only 3 full-symmetric $A_{1g}$ modes are visible
(Fig.~\ref{lgraman}), with relevant parameters given in
Table~\ref{ramantab}.

%-------------
%-------------

%-------------
%-------------

\begin{table}[h]
\caption{\label{ramantab}Raman gain of undoped LiSGaF.} \vspace*{1 cm}
\begin{center} 
\begin{tabular}{|c|c|c|c|}
\hline%
&&&\\%
 Frequency $\Omega_j/2\pi$ & Raman gain $g^s_j$ & Width (FWHM) &
$T_j$\\%
cm$^{-1}$ & cm/GW & cm$^{-1}$ & ps \\%
\hline%
&&&\\
 230 & $0.014 \pm 0.005$ & $9 \pm 3      $ & $1.2 \pm 0.4   $  \\%
 349 & $0.021 \pm 0.006$ & $14 \pm 2     $ & $0.7 \pm 0.1 $  \\%
 551 & $0.32  \pm 0.05$  & $12.5 \pm 0.6 $ & $0.86 \pm 0.05 $  \\%
&&&\\ \hline
\end{tabular}
\end{center}
\end{table}

%%------------------------------------------------------------
\section{MODEL}

There exist different approaches to modeling of ultrashort pulse generation
in solid-state laser, which are based on soliton or fluctuation models. The
soliton approach can be applied only the distributed laser model (where the
dispersion and nonlinearity are implied to be evenly distributed over the
round-trip and act simultaneously) but allows to build comparatively simple
analytical description thus promoting easy interpretation of results. We also
based our calculations on the distributed laser model but the results were
tested by simulations on the basis of discrete-element scheme corresponding
to Fig.~\ref{setup}. To overcome the limitations of the soliton approach we
used numerical simulations allowing to account for the high-order dispersion,
the laser field reabsorption, the complicated spectral profiles of the gain
and output coupler transmission, and the Raman scattering within the active
medium.

The modelocking is described by a fast absorber-like action of  Kerr-lensing
in the active medium in the form of a nonlinear transmission-operator
    $exp\left[-\frac{\gamma }{1+\sigma \left| a(z,t)\right| ^{2}}%
\right]$ ~\cite{APMKLM}, where $\gamma $ is the modulation depth (KLM loss),
which is set by the cavity arrangement, $\sigma $ is the inverse loss
saturation intensity, $a$ is the field depending on local time $t$ and
longitudinal
coordinate $z$ ($|a|^{2}$ has the meaning of the field intensity). Parameters $\gamma $ and $%
\sigma $ are controlled by changing the cavity configuration, which is a
common procedure for Kerr-lens mode-locking experiment.

Another fundamental factor in our model is the presence of high-order
dispersions due to the active medium, the dispersion compensator, the output
coupler, and high-reflective mirrors. The corresponding experimental
characteristics are shown in Fig.~\ref{DispFig}. For the numerical
calculations, the data were represented by the eighth-order polynomial
approximation.  The action of dispersion can thus be presented in the
following form:

\begin{equation}
a(z,t)=\int\limits_{-\infty }^{\infty }a(z,t^{^\prime })G(t-t^\prime
)dt^\prime , \label{adef}
\end{equation}
\begin{equation}
G(t-t^\prime )= \frac{1}{2\pi }\int\limits_{-\infty }^{\infty} \exp{ \left(
-i \sum_{j=2}^8 \frac{1}{j!}D_{j}(\omega-\omega_0)^{j}-\left( t-t^\prime
\right) \omega \right) }d\omega, \label{Green}%
\end{equation}%

\noindent%
where $\omega$ is the frequency, $G(t-t^\prime)$ is the Green's function
depending on the dispersion coefficients $D_{j}$ up to eighth order of $j$.

The next important factor in our model is the gain saturation that was
described on the basis of quasi-two level scheme of the active medium
operation ~\cite{LUL}:

\begin{equation}
\label{gsat}%
\frac{\partial \alpha }{\partial t}=%
\frac{I_{p}\sigma _{14}}{h\nu }\left(\alpha _{\max }-\alpha \right)
-\frac{\left|a\right| ^{2}\sigma _{32}}{h\nu }\alpha -\frac{\alpha }{T_{r}}%
\end{equation}

\noindent%
where $I_{p}$ is the pump intensity, $\nu $ is the pump frequency, $\sigma
_{14}$ and $\sigma _{32}$ are the loss and the gain cross-sections,
respectively, $T_{r}=85\;\mu$s is the gain relaxation time. If the pulse
duration is much less then cavity round-trip time  $T_{cav}=14\;$ns, then
this equation can be replaced by the following one:

\begin{equation}
\label{gsatmod}%
\frac{\partial \alpha }{\partial z}=P\left( \alpha _{\max }-\alpha \right) -%
\frac{E}{E_{s}}\alpha -\frac{T_{cav}}{T_{r}}\alpha%
\end{equation}

\noindent%
where $P=(I_{p}\sigma _{14}/h\nu)T_{cav}$ is the dimensionless pump intensity,
$z$ is the dimensionless longitudinal coordinate, i. e. the number of the
cavity round-trips. $E_{s}=h\nu/\sigma _{32}$ is the gain saturation energy
flux, $E$ is the full pulse energy flux.

\noindent
Self-phase modulation in the active medium was represented by nonlinear
``transmission'' operator $exp\left[-i\beta \left| a(z,t)\right|
^{2}\right]$, where $\beta ={2\pi n_{2}x}/{\lambda
n}=3.4\;\mathrm{cm^{2}}/\mathrm{TW}$ is the self-phase modulation
coefficient. Here $n$ and $n_{2}$ are the linear and nonlinear coefficients
of refraction, respectively, $\lambda$ is the central wavelength
corresponding in our case to the gain band maximum, $x=8\;$mm is twice the
length of the active crystal.

Finally, we consider the stimulated Raman scattering within the active
medium. Following Ref. ~\cite{JosaHaus}, where Raman scattering contribution
was calculated analytically on the basis of the soliton model, we supplement
the model with the following equations:

\label{ramanset}
\begin{eqnarray}
\frac{\partial a_{s}}{\partial \varsigma } & = & i\sum\limits_{j=1}^{3}Q_{j}^{%
\ast }a_{p}, \\
\frac{\partial a_{p}}{\partial \varsigma } & = & i\sum%
\limits_{j=1}^{3}Q_{j}a_{s},  \\
\frac{\partial ^{2}Q_{j}}{\partial t^{2}}+\frac{2}{T_{j}}\frac{\partial
Q_{j}}{\partial t}+\Omega _{j}^{2}Q_{j} & = & \mu _{j}a_{p}a_{s}^{\ast },%
\end{eqnarray}%

\noindent%
where $\varsigma $\ is the longitudinal coordinate (pulse propagation axis)
inside the active medium, $a_{p,s}$ are the amplitudes of the ``pump'' and
the ``Stokes'' components within generation spectrum, $\Omega _{j}$ are the
phonon resonance frequencies ($j=1,2,3$ corresponding to the three
Raman-active phonon resonances in LiSGaF, see Fig.~\ref{lgraman}).
$T_{j}$ are the inverse bandwidths of Raman lines, $\mu _{j}$=$%
g_{j}^{s}\Omega _{j}/T_{j}$ are the coupling parameters for Raman gain
coefficients $g_{j}^{s}$.

Solving third equation of the system, we obtain the steady-state phonon
amplitude for the fixed pump and the Stokes components with corresponding
frequencies $\omega _{p}$\ and $\omega _{s}$:

\begin{equation}
\label{five}%
 Q_{j}=\frac{\mu _{j}a_{p}a_{s}^{\ast }}{\Omega
_{j}^{2}-\frac{2i\left( \omega _{p}-\omega _{s}\right) }{T_{j}}-\left( \omega
_{p}-\omega _{s}\right) ^{2}}\approx \frac{\mu _{j}a_{p}a_{s}^{\ast
}}{2\Omega
_{j}\left( \Omega _{j}-\left( \omega _{p}-\omega _{s}\right) \right) -\frac{%
2i\Omega _{j}}{T_{j}}}.
\end{equation}

The validity of the last approximate expression follows from the fact that
the Raman lines are narrow in comparison with the pulse spectrum. With regard
to the contribution of all spectral components of the pulse to the phonon
amplitude, the equation for the Stokes and the pump fields in the frequency
domain can be written as:

\label{six}
\begin{eqnarray}
\frac{\partial a_{s}}{\partial \varsigma } &=&ia_{s}\sum\limits_{j=1}^{3}\mu
_{j}\sum_{k}\frac{\left| a_{p,k}\right| ^{2}}{2\Omega _{j}(\Omega
_{j}-(\omega _{p,k}-\omega _{s}))+\frac{2i\Omega _{j}}{T_{j}}} \\
\frac{\partial a_{p}}{\partial \varsigma } &=&ia_{p}\sum\limits_{j=1}^{3}\mu
_{j}\sum_{k}\frac{\left| a_{s,k}\right| ^{2}}{2\Omega _{j}(\Omega
_{j}-(\omega _{p}-\omega _{s,k}))-\frac{2i\Omega _{j}}{T_{j}}}
\end{eqnarray}%

\noindent%
where $k\ $is the index of the field's spectral (i.e. Fourier) component
(in the simulations we considered $2^{13}$ components).
Since the Raman lines are narrow, the field variation within these lines is
negligible, i.e. $a_{s,p}$ are constant and can be taken out of the second
summation. Then the summation can be executed explicitly by transition to the
integral, resulting in

\label{parametric}
\begin{eqnarray}
\frac{\partial a_{s}}{\partial \varsigma } &=&\frac{\pi }{4}%
\sum\limits_{j=1}^{3}a_{s}g_{j}^{s}\left| a_{p}\right| ^{2}, \\
\frac{\partial a_{p}}{\partial \varsigma } &=&-\frac{\pi }{4}%
\sum\limits_{j=1}^{3}a_{p}g_{j}^{s}\left| a_{s}\right| ^{2}.
\end{eqnarray}

It should be noted, that in these equations $\omega _{p}-\omega
_{s}=\Omega_{j}$ and there is the pair-wise interaction of the spectral
components within the wide enough generation spectrum.

There are two main mechanisms of the generation of the initial seed at the
Stokes frequency for Eqs. \eqref{parametric}. The first one is the
spontaneous Raman scattering with the increments of the scattered intensity
growth

\begin{equation}
\chi _{j}=\frac{\omega _{p}\omega _{s}^{2}n_{s}^{2}\hbar g_{j}^{s}}{\pi c^{2}%
\left[ 1-\exp \left( -\frac{\hbar \Omega _{j}}{k_{B}T}\right) \right] },
\end{equation}

\noindent%
where $n_{s}$ is the index of refraction at Stokes frequency, $T$ is the
temperature, and $k_{B}$ is the Boltzmann's constant ~\cite{NLO}.
With this seed signal, the stimulated Raman scattering results in appearance
and growth of spectral replicas of the main oscillation pulse, shifted to the
lower frequencies by the Raman frequencies $\Omega_j$.

More significant source for Stokes component's amplification, however, is the
broad-band pulse field itself. When the pulse pulse spectrum is wide enough
to become comparable with the Raman frequency shift, the lower-frequency part
of the spectrum can play a role of the Stokes component seed with respect to
the higher-frequency part of the spectrum. The stimulated Raman scattering
then transfers the energy form the higher-frequency components to the
lower-frequency ones, resulting in the continuous red-shift of the pulse
spectrum as a whole. As the field amplitude of the laser pulse significantly
exceeds the spontaneous seed, the second mechanism strongly dominates over
the first one. However, we included both mechanisms in our model, because
their influence on the pulse spectrum is quite different.

Later on it is convenient to normalize the time to the inverse gain bandwidth
$t_{g}=2.25\;$fs and the intensity to $\beta ^{-1}$, resulting in the
normalization of the field energy to $(\beta t_{g})^{-1}$. As already pointed
out, we analyzed the described above model in two ways: on the basis of
distributed and discrete-element approaches. In the framework of the
distributed model, we didn't consider the propagation through the individual
laser element and supposed that the pulse envelope is formed by the overall
net-dispersion in the cavity. As result we have a split-step scheme
describing ultrashort pulse propagation from $z$ to $z+1$ transits:

\label{splitstep}
\begin{equation}
a(z^\prime ,t)=\int\limits_{-\infty }^{\infty }...\int\limits_{-\infty }^{\infty }a(z,t^\prime
)C(t-t^\prime )L\left( t^\prime -t^{\prime\prime} \right) 
A\left( t^{\prime\prime} -t^{\prime\prime\prime} \right) G\left( t^{\prime
\prime \prime} -t^{\prime \prime \prime \prime} \right) dt^{\prime}
dt^{\prime \prime} dt^{\prime \prime \prime} dt^{\prime \prime \prime \prime}
, \end{equation}
\begin{equation}
 a(z+1,t)=a\left(
z^\prime ,t\right) \exp \left( -\frac{\gamma }{1+\sigma \left| a(z^\prime
,t)\right| ^{2}}-i\left| a(z^\prime ,t)\right| ^{2}\right) ,
\end{equation}
\begin{equation}
\alpha (z+1)=\alpha (z)\exp \left( -\tau \int\limits_{-\infty }^{\infty
}\left| a(z,t^\prime )\right| ^{2}dt^\prime -{T_{cav}}/{T_{r}}-P\right) +
\frac{P\alpha _{\max }\left( 1-\exp \left( -{T_{cav}}/{T_{r}}-P\right)
\right) }{P+{T_{cav}}/{T_{r}}},%
\end{equation}

\noindent%
where $\tau ={t_{g}}/({E_{s}\beta })=$ 0.00079 is the normalized gain
saturation parameter. The Green's functions $A,\;L,\;C$\ describe spectral
bands of gain, reabsorption and output loss, respectively (note, that the
dispersion is already is included in \textit{G}):

\label{ALC}
\begin{eqnarray}
A(t-t^\prime ) &=&\frac{1}{2\pi }\int\limits_{-\infty }^{\infty }(1+\alpha
\left( z\right) )\Phi _{\alpha }\left( \omega \right) \exp \left[ i\omega
\left( t-t^\prime \right) \right] d\omega , \\ L\left( t-t^\prime \right)
&=&\frac{1}{2\pi }\int\limits_{-\infty }^{\infty }\Phi _{r}\left(
\omega \right) \exp \left[ i\omega \left( t-t^\prime \right) \right] d\omega
, \\ C(t-t^\prime ) &=&\frac{1}{2\pi }\int\limits_{-\infty }^{\infty
}\Phi _{out}\left( \omega \right) \exp \left[ i\omega \left(
t-t^\prime \right) \right] d\omega ,
\end{eqnarray}

\noindent%
Here, $\Phi _{\alpha ,\;r,\;out}$\ are the ``form-factors'' describing
spectral profiles of the gain, reabsorption and the output coupler bands,
which resulted from the functional approximation of the experimental data
(Fig.~\ref{LossGain}).

The system \eqref{splitstep} has to be completed by the system
\eqref{parametric} and the result can be solved on the basis of numerical
simulation in Fourier domain and split-step method for nonlinear propagation.

The discrete-element approach is based on the element-to-element
simulation of the pulse propagation through the cavity on every round-trip,
following the laser scheme in Fig.~\ref{setup}. The nonlinear and spectral
characteristics of each laser element are considered separately. Further
refinement of the simulation is achieved by considering the pulse propagation
through active medium by splitting it into five slices and applying the
split-step procedure  to each slice consecutively.
%%-----------------------------------------------------------
\section{Discussion}

Our simulations are aimed at investigation of the influence of the different
factors on spectral characteristics of the ultrashort pulses. Therefore, to
simplify the interpretation, we will first consider  the high-order
dispersion action without Raman scattering and without reabsorption; then
reabsorption and Raman scattering will be taken into consideration without
high-order dispersion action; and finally, the join action of all factors
will be analyzed. To conclude, we will also compare the  obtained results
with the calculation based on the discreet-element model.

\subsection{High-order dispersion action}

As it was discussed in ~\cite{Akhmanov}, small contribution of third-order
dispersion to characteristics of Schr\"odinger soliton does not cause the
frequency shift, but introduces additional field time delay on the cavity
round-trip. However, the influence of the linear and nonlinear dissipative
terms in Eqs.~\eqref{splitstep} can destroy the soliton character of the
pulse, in particular, it can add the chirp. The latter, as it was shown in
~\cite{NewMex}, results in the frequency shift of the pulse spectrum in the
non-steady-state regime. Additionally, in the real-world laser systems the
contribution of the high-order dispersions, as a rule, lies beyond the bounds
of the perturbation theory validity.

\begin{figure}
%>>>> following adds vertical space needed for figure; 
%  uncomment if figure is to be pasted into manuscript
   \begin{center}
   \begin{tabular}{c}
   \psfig{figure=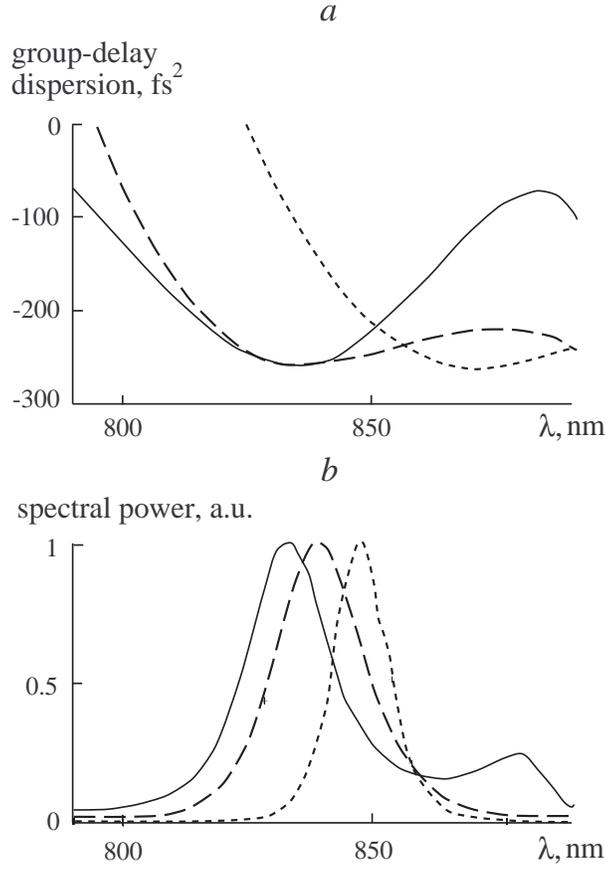,height=12cm} 
   \end{tabular}
   \end{center}
   \caption[example] 
   { \label{Dcurves}	  
The dependence of dispersion (\textit{a}) and generation
spectra (\textit{b}) on wavelength. \textit{P} = 3.2$\times10^{-4}$, $\sigma$
= 1, $\gamma$ = 0.05, pulse energy \textit{E} is 20 nJ. Pulse durations
$t_p$: 27 (solid curve), 38 (dash), 36 fs (dot).} 
   \end{figure}

The typical net-dispersion curves corresponding to the experiment with the
chirped mirrors are presented in Fig.~\ref{Dcurves}, \textit{a}. Over the
full spectral region of generation the pulse undergoes the non-negligible
influence of the dispersion up to the eighth-order (see Eq.~\eqref{Green}).
As a rule, there is the closed spectral window of dispersion, which is ``optimal'' for
steady-state pulse generation. The contribution of the high-order dispersion
terms can shift this window. In Fig.~\ref{Dcurves}, \textit{a}, this shift
corresponds to the red-shift of the positive net-dispersion branches
(transition from solid to dash and dot curves). The resulting output spectra
obtained from the distributed model are shown in Fig.~\ref{Dcurves},
\textit{b}. The net-dispersion shift is accompanied by the red-shift of the
pulse spectrum (transition from solid to dash and dot curves). Besides this
effect there is the possibility of the essential spectral profile distortion
(solid curve) and side-band generation (side-bands lie outside the of shown
region) ~\cite{ICONO}.

%-------------
%-------------

\begin{figure}
   \begin{center}
   \begin{tabular}{c}
   \psfig{figure=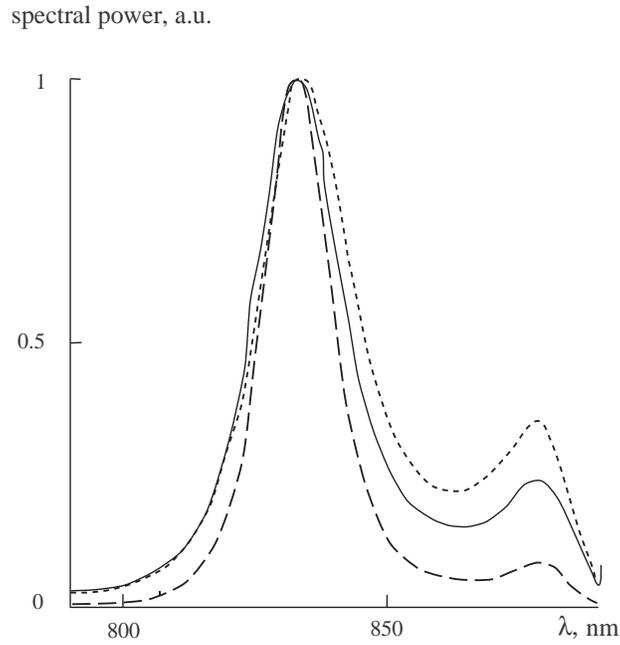,height=9cm} 
   \end{tabular}
   \end{center}
   \caption[example] 
   { \label{spectra}	  
The dependence of generation spectra on
wavelength  in the presence of high-order dispersion in the
distributed (solid and dash curves) and the discrete-element (dot)
models. $\sigma$ = 1, $\gamma$ = 0.05. For the corresponding pulse
parameters see Table~\ref{tabfig}.} 
   \end{figure}

However, we cannot consider this shift as the cause of the experimentally
observed effect because the dispersion shift has the linear nature, i. e.
there is no obvious dependence of this shift on the field energy. This is
demonstrated by Fig.~\ref{spectra}, where the pulse energy variation due to
the pump variation changes the spectral profile, but does not cause any
noticeable frequency shift (compare solid and dash curves in this figure).

%-------------
%-------------

The obtained results demonstrate that the self-frequency shift cannot be
caused by the non-dissipative factors. As the pulse duration is too large for
the nonlinear dispersion to play any significant role, we will concentrate on
spectrally-dependent losses and Raman effect.

\subsection{Output loss and reabsorption in gain medium}

\begin{figure}
   \begin{center}
   \begin{tabular}{c}
   \psfig{figure=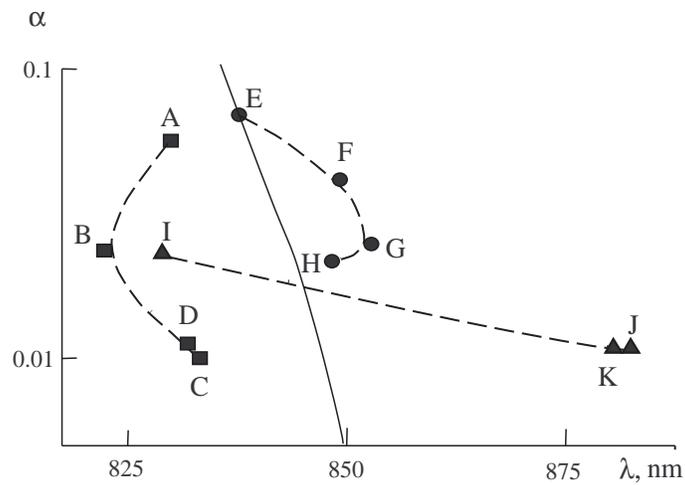,height=7cm} 
   \end{tabular}
   \end{center}
   \caption[example] 
   { \label{shifts}	  
Pulse central wavelength as a function of the
saturated gain coefficient. Solid line: Net-gain maximum; $ABCD$
spectrum peak in the case of gain saturation without reabsorption
and Raman scattering in the active medium; $EFGH$ contribution of
the reabsorption; $IJK$ contribution of Raman scattering. Points
correspond to Table~\ref{energies}.} 
   \end{figure}

As mentioned above, dependence of the frequency shift on pulse energy implies
the involvement of some nonlinear mechanism. Since in the real-world systems
the gain band does not coincide with filtering band (output mirror in our
case) and the reabsorption band, the spectral position of the net-gain
maximum changes with the gain value. The latter is defined by the pump and by
the pulse energy (see Eqs.~\eqref{splitstep}): pulses with higher energy
experience lower gain due to the multi-pass saturation. The dependence of the
net-gain maximum on the saturated steady-state gain coefficient $\alpha$ is
shown in Fig.~\ref{shifts} by the solid curve. This curve was obtained from
the numerical analysis of the measured spectrum of the  output coupler,
intracavity loss and gain profiles. As we can see from this figure, the
behavior of the net-gain maximum corresponds to the Stokes frequency shift
with the pulse energy growth, due to the gain coefficient decrease. However,
the magnitude of this shift ($\sim 10\;$nm) is not sufficient to explain the
experimental values (up to 50 nm). Besides that, the location of the pulse
spectrum in general does not coincide with the net-gain maximum.

%-------------
%-------------

\begin{table}[h]
\caption{\label{tabfig} Normalized pump power, pulse duration and
energy for Fig.~\ref{spectra}}
\bigskip
\begin{center}
\begin{tabular}{|c|c|c|c|c|c|c|c|c|c|}
\hline Line in Fig.~\ref{spectra}&\textit{P$\times10^4$}& $t_p$,
fs& \textit{E}, nJ
\\
\hline solid& 3.2& 27& 20
\\
\hline dash& 2& 30& 12
\\
\hline dot& 3& 25& 16
\\
\hline
\end{tabular}
\end{center}
\end{table}

\bigskip

\begin{table}[h]
\caption{\label{energies}Ultrashort pulse energies and durations
for Fig.~\ref{shifts}}
\bigskip
\begin{center}
\begin{tabular}{|c|c|c|c|c|c|c|c|c|c|c|c|c|}
\hline Points& \textit{A}& \textit{B}& \textit{C}& \textit{D}&
\textit{E}& \textit{F}& \textit{G}& \textit{H}& \textit{I}&
\textit{J}& \textit{K}\\%
\hline Energy, nJ& 16& 27& 47& 54& 20& 38& 52& 119& 33& 78& 103\\%
\hline Pulse width, fs& 45& 28& 20& 23& 28& 20& 18& 37& 25& 14&
19\\%
\hline
\end{tabular}
\end{center}
\end{table}
%-------------

To check the last thesis we performed numerical simulations on the
basis of the described above model in the presence of dissipative
factors only. The results are presented in Fig.~\ref{shifts} by
the dashed lines $ABCD$ and $EFGH$. The letters alongside of
squares, circles and triangles denote the intracavity pulse energy
obtained from the numerical simulation (see Table~\ref{energies}).
The pulse energy variation results from the change of modulation
parameter $\sigma$, i.e. due to the change of the cavity
configuration.

The curve $ABCD$ illustrates the case when there is no
reabsorption and Raman scattering in the active medium. The
dependence of frequency shift on pulse energy is not monotonous:
at the minimum pulse width (15 fs in our case) there is no Stokes
frequency shift, contradicting the experimental data. Moreover,
the maximal shift (of about 6 nm) is by an order of magnitude less
than that observed in the experiment. It is clear, that gain band
asymmetry and output loss alone cannot produce large
self-frequency shift in Cr:LiSGaF laser.

Stronger contribution is provided by the reabsorption in the
active medium (curve $EFGH$ on Fig.~\ref{shifts}). For small
energies the simulation result (point \textit{E} on
Fig.~\ref{shifts}) coincides with the prediction made on the basis
of elementary model of net-gain maximum shift (solid curve). The
rise of the pulse energy increases the shift (up to 25 nm in our
case), much more than in the previous case, although still by a
factor of two less than in the experiment.

Note, that after certain pulse energy the spectral shift is again
decreasing. This is caused by the increase of the pulse duration
for the large energies (point \textit{H} on Fig.~\ref{shifts}),
due to the nonlinear loss saturation, resulting in reduced
spectrum width. The laser approaches the condition of the  cw
operation, described by solid curve (although the conditions of
pulsed gain saturation strongly differ from those in cw-regime).

\subsection{Raman scattering}

As pointed out in the previous section, the net-gain shift model
fails to provide correct description of the experimental data by a
factor of two. However, taking the stimulated Raman scattering
into account allows to obtain large frequency shifts, increasing
with pulse energy growth. In Fig.~\ref{shifts} the curve $IJK$
demonstrates the Raman scattering action in the absence of
reabsorption effect. As the simulation demonstrates, the red
components originate from the amplified Raman signal, which pulls
the whole spectrum over the long wavelength limit at the given
pump power, defined by the spectral filtering.

Note the pronounced threshold-like character of the effect. For
the small energies (point $I$) the spectral shift is negligible,
but the energy growth causes very strong shift (60 nm in our case)
in good agreement with experimental results
~\cite{14fs,sub20fs,ICONO} and with analytical prediction
~\cite{JosaHaus}. Since the gain saturation does not play important
role in this case, the dependence of frequency shift on gain
coefficient is insignificant.

As already mentioned above, the main contribution to the stimulated Raman
process comes from the energy transfer from the blue part of the pulse
spectrum (pump) to the red one (Stokes). The efficiency of the stimulated
Raman scattering is therefore defined by the product of intensities at pump
and  Stokes frequencies. The separation between Stokes and pump components is
fixed, it is equal to the Raman line frequency $\Omega_j$. Therefore,
decreasing the pulse spectrum width strongly suppresses the effect and
reduces self-frequency shift. Assuming that the pulse spectrum has
exponential fall-off to the blue and red sides, we see that the dependence of
the Raman shift on the pulse spectrum width should be asymptotically
exponential at long pulse durations. This is also the result of the
analytical theory in Ref~\cite{JosaHaus}, where sech$^2$ pulse shape has been
assumed.

As the Raman effect strongly depends on the pulse intensity and
Raman gain, it should be especially pronounced in low-gain lasers
working with low output coupling and high intracavity pulse
energy, such as Cr:LiSGaF, Cr:LiSAF, Cr:YAG. These materials also
possess strong and broad Raman lines ~\cite{TOPS98}. Large
power-dependent red-shift in Cr:LiSGaF and Cr:LiSAF is well
documented ~\cite{14fs,12fs,sub20fs}. In femtosecond Cr:YAG lasers,
femtosecond pulse spectrum is also always red-shifted with respect
to the cw wavelength in the same resonator ~\cite{CrYAG}.

\subsection{Discrete-element model}

Finally, we can compare the simulation results in case of
distributed and discrete-element models. In the case of the
high-order dispersion action the transition to the
distributed-element model does not significantly change the
spectral characteristics of the pulse (the dot curve in
Fig.~\ref{Dcurves}, \textit{b}). The long-wavelength ``shoulder''
of the spectrum in the case of the net dispersion corresponding to
solid curve in Fig.~\ref{Dcurves} \textit{a} is somewhat stronger
than in the distributed model. This is because this ``shoulder''
results from local dispersion maximum due to the chirped mirror
(CM$_1$ on Fig.~\ref{setup}), which is the closest to the output
mirror. Additionally, the self-phase modulation contribution is
found to be slightly higher in comparison to the distributed
model. However, all these changes are not qualitative. We found
also that the transformation of the spectrum on each laser
elements is small ensuring the validity of the distributed model.
It should also be noted, that the transition from the distributed
to the discrete model slightly increases the contribution of the
Raman scattering, seen by the lower threshold energy.

\section{Conclusion}
Using the numerical simulations performed in the framework of one-dimensional
distributed and discrete-element models, we analysed the spectral
characteristics of a cw Kerr-lens mode-locked Cr: LiSGaF-laser. The two main
factors causing the ultrashort pulse self-frequency shift have been
established: the nonlinear shift of the net-gain band due to the gain
saturation in the presence of reabsorption in the active medium and the
stimulated Raman scattering. The first effect is essential for comparatively
small pulse energies and produces wavelength shifts up to 30 nm. The Raman
scattering occurs as a result of pulse energy growth and causes the large
(over 50 nm) red shift. The contribution of the high-order dispersion (up to
the eighth order), gain-band asymmetry and spectral characteristics of output
coupler were estimated as well.  However, their effect on the pulse central
frequency is much smaller than that of the stimulated Raman scattering, which
is the main cause of spectral red-shift in Kerr-lens modelocked laser. The
shift values obtained from the numerical simulations are in good agreement
with experimental data.

Analytical part of the calculations is presented on \textit{http://www.geocities.com/optomaplev}

\section*{ACKNOWLEDGMENTS} 
This work was supported by Austrian National Science Fund Projects T-64, P14704-PHY and
Austrian National Bank Project 7913.
V.L. Kalashnikov gratefully acknowledges support from the Austrian Science Fund (FWF), 
Project M611.

  \end{document}